\documentstyle[graphicx,psfrag,amssymb,prl,twocolumn,aps]{revtex}

\begin{document}
\draft
\twocolumn[\hsize\textwidth\columnwidth\hsize\csname @twocolumnfalse\endcsname
\title{
Specific heat of the Kelvin modes 
in  
low temperature superfluid 
turbulence} 
\author{Uwe R. Fischer} 
\address{
Eberhard-Karls-Universit\"at T\"ubingen, 
Institut f\"ur Theoretische Physik \\
Auf der Morgenstelle 14, D-72076 T\"ubingen, Germany}

\date{\today}
\maketitle


\begin{abstract}                
It is pointed out that 
the specific heat of helical vortex line excitations,  
in low temperature superfluid turbulence experiments 
carried out in helium II, can be of the same order as the specific heat 
of the phononic quasiparticles. 
The ratio of Kelvin mode and phonon specific heats scales 
with $L_0 \, T^{-5/2}$, where $L_0$ represents the smoothed 
line length per volume within the
vortex tangle, such that the contribution of the vortex mode specific heat
should 
be 
observable for 
$L_0 = 10^6 \cdots 10^8 \, {\rm cm}^{-2}$,  
and at temperatures which are of order $1\cdots 10$ mK.
\end{abstract}
\


\

] \narrowtext

The problem of superfluid turbulence has seen a renewed  
upsurge of interest, largely due to the fact that turbulence
in superfluids at temperatures of order one Kelvin 
\cite{smith93,maurer,stalp}, has proved to be similar 
to turbulence in normal fluids \cite{frischbook}. 
Of particular interest should be the range of  
lower temperatures, of order a few mK, where the normal fluid density 
is negligible, and a particularly interesting behaviour of 
superfluid turbulence in a very pure form should be 
observable \cite{vinen2000,hendry2000}.  
Since the pioneering work of Vinen \cite{vinen57}, 
the dynamics of superfluid turbulence, especially its generation and 
decay, is to a large extent 
still not understood. 
More recent efforts in this direction, investigating, {\it e.g.}, 
the energy spectrum and reconnection events 
in the tangle, can
be found in \cite{noreabidbrachet97,tsubotaPRB00}. An interesting thermodynamic
approach to directly observable effects of dissipation in low temperature 
superfluid turbulence has been proposed by Samuels and Barenghi 
\cite{samuelsbarenghi98}.  
The effect to be observed, according to this proposal, 
consists in the conversion of (incompressional) kinetic energy 
of the vortex tangle into compressional energy, which then heats the sample. 
In that work, it has been assumed that the specific heat of the turbulent 
helium II sample is given solely by phononic excitations.  
Here, it will be shown that, at temperatures of the 
order a few mK, the total specific heat of line excitations 
in the vortex tangle 
is of the same order as the 
specific heat of phonons, integrated over the volume of the tangle. 
This will be true for large, yet experimentally feasible, line densities 
in the tangle. Though vortex lines fill out a very small fraction of the 
volume of the turbulent helium sample, the contribution of their oscillations 
to the specific heat should, as will now be shown, 
 be discernible from the phonon specific heat at 
low temperatures.  
   

The total specific heat of the helium II sample is assumed to be composed 
of the specific heat of the elementary {nontopological} 
excitations which constitute the heat bath, living in the three dimensions
of the sample, and the specific heat of the excitations 
propagating along the topological line vortices, 
which live in one spatial dimension. 
Below temperatures of a fraction of one Kelvin, only phonons 
contribute to the specific heat of the nontopological 
excitations, the contribution of the gapful excitations (rotons) 
being exponentially suppressed. 
The specific heat of sound excitations in a volume $\cal V$ is 
\begin{eqnarray}
\frac{C_{\rm ph}}{k_B} & = & 
\frac{2\pi^2}{15} {\cal V} \left( \frac{k_B T}{\hbar c_s}\right)^3
\,,  \label{Cph}
\end{eqnarray} 
where $c_s$ is the speed of sound. 
The specific heat of the oscillation modes of an isolated 
vortex ring of radius $r_0$ 
may generally be written in the form 
\begin{eqnarray}
\frac{C_{\rm L}}{k_B} 
\label{CL}
& = & \frac{2\pi r_0}{\xi_c}\, f \left(\frac{k_B T}{\hbar\omega_c} \right) 
\,. 
\end{eqnarray}
The nondimensional function $f$ depends on the ratio of the 
temperature and the `cyclotron' energy of the vortex core,   
$k_B T/\hbar\omega_c$, 
characterizing the separation 
between thermal and core energy physics \cite{BDVPhysFluids85}.
In helium II, $\hbar\omega_c \sim 1 \cdots 10$ K, so that the 
parameter
$k_B T / \hbar\omega_c \sim 10^{-3}\cdots 10^{-4}$ for temperatures in the
mK range.     
The cyclotron frequency of the core is defined by  
$
\omega_c={\Gamma}/({\pi (\xi_c/2)^2})\,, 
$ 
with $\xi_c$ the core diameter and $\Gamma = h /m$ the quantum 
of velocity circulation. For a line of arbitrary global shape, $2\pi r_0$ 
in (\ref{CL}) is replaced by the smoothed 
arc length of vortex line, {\it i.e.}, the length of line 
with the Kelvin disturbances subtracted \cite{vinen2000}. 
This undisturbed, smoothed  length of line plays the role of ``volume'' 
in the specific heat of the Kelvin modes. 
The smoothed line density of vortices 
is designated $L_0=L_0[{\rm cm}^{-2}]$, 
so that we may assign $2\pi r_0 = L_0 {\cal V}$, as an (average)  
filament length entering the specific heat (\ref{CL}). 
The line density $L_0$, around which there are Kelvin 
fluctuations, is to be 
distinguished  from the full line density $L$,  
which derives from the total (incompressible) 
kinetic energy, with the contribution of the Kelvin waves included.
They differ by a factor logarithmically dependent on the product 
of the largest Kelvin wave number excited and the 
average vortex element 
distance $l=L_0^{-1/2}$ (the smallest possible 
Kelvin wave number is $l^{-1}$) \cite{vinen2000}, so that $L_0$ can be 
about an order of magnitude less than $L$.  

The function $f$ in (\ref{CL}) depends on the spectrum of waves on the 
vortex line.  
For wavelengths much larger than the core diameter, 
a single ring vortex has the Kelvin spectrum 
\begin{eqnarray}
\omega_{\rm K} & = &
\frac{\Gamma n^2}{4\pi r_0^2} \ln [{n_c^*}/{n}]\,, \label{Kelvin}
\end{eqnarray}
where $n$ is the number of waves on the circumference (the mode number),
so that the wave number 
$k=n/r_0$.  
The ultraviolet cutoff mode number 
is for a vortex ring  $n_c^* = O(2\pi r_0/\xi_c)$; 
for a filament of more general shape, the cutoff wave number
$k_c=O(\xi_c^{-1})$, and $n_c^*$ is of order the total length 
of a filament, divided by $\xi_c$. 

If vortex filaments approach each 
other to within a distance corresponding to the inverse Kelvin wave vector 
under consideration, so that $kl\lesssim 1$, 
the Kelvin spectrum will be changed.  
In the expression (\ref{Kelvincontr}) 
for the specific heat of the oscillations  
presented below, we will 
assume that vortices remain, at least on average, 
sufficiently well separated to retain the validity of the Kelvin 
spectrum (\ref{Kelvin}) for most of the 
excitations on the filaments.  
Deviations from that spectrum 
are then too small to be thermodynamically relevant for
the specific heat 
of the oscillations, and in particular its temperature dependence. 
Conversely, if there are any measurable deviations from the temperature 
dependence of the specific heat of Kelvin waves,  
we will have an indicator 
that the Kelvin spectrum has changed due to the presence of a 
very dense vortex tangle.  
Strictly spoken, these conclusions can 
only claim validity if the vortex system is in thermodynamic 
equilibrium with itself and the surrounding fluid.
This is the case for a vortex array, generated by a constant rotation rate
$\Omega$ of the superfluid, with resulting 
densities $L_0 = 2\Omega/\Gamma= 2\cdot 10^3 {\rm cm}^{-2}{\rm sec}\,
\,\Omega$. 
The rotation rates to achieve the line densities of order
$L_0 = 10^6 \cdots 10^8 $ cm$^{-2}$, discussed below, are thus in the range 
$10^3 \cdots 10^5 $ rad/sec, {\it i.e.}, of an order of magnitude 
reached in centrifuges only, so that this appears difficult to realize
in experimental low temperature practice.  
In general, the turbulent vortex tangle will not be in a
(local) thermodynamic equilibrium state. 
We may, however, expect that the contribution 
of line oscillations in the tangle to the internal energy 
will still be proportional to the line density $L_0$. 
Furthermore, any deviation from the predicted 
dependence of the line oscillation contribution on temperature (for
line densities which are not too high, so that the Kelvin spectrum 
remains valid),  
will give a measure of how {\em far}
the turbulent tangle is away from thermodynamic equilibrium.

The one-dimensional density of states for the Kelvin modes, within 
logarithmic accuracy, may be written in the form  
\begin{equation}
N_{\rm K}(E)\simeq \frac{\sqrt{\pi}\,r_0}{\sqrt{\hbar \Gamma \ln[n_c^*]}}\, E^{-1/2}
\,.
\end{equation}  
For rings large compared  to $\xi_c$, 
in the continuum approximation of closely spaced excitation 
frequencies, this expression for the density of states 
then enables the calculation of the specific heat 
of the Kelvin modes according to 
\begin{equation}
C_{\rm L}= L_0{\cal V} \frac{\partial}{\partial T}\left( 
\int_0^\infty\!\! dE\, N_{\rm K}(E)\, \frac {E}{\exp[E/k_B T]-1}\right).
\end{equation}
There results the following expression for the 
dimensionless function in (\ref{CL})
\begin{equation}  
f_{\rm K}= \frac{3\zeta(\frac32)}{4\sqrt{\pi \ln[n^*_{c}]}}
\,\sqrt{\frac{k_B T}{\hbar\omega_c}}\,.
\end{equation}
More explicitly, 
we may write for the specific heat of the Kelvin modes 
\begin{eqnarray}
\frac{C_{\rm L}}{k_B} 
& \simeq & 
{\cal V} L_0\, \frac{3\zeta(\frac32)}{8}\,
\sqrt{\frac{k_B T}{\hbar \Gamma \ln [n_c^*]}}
\nonumber\\
& = & 
3.54
\cdot 10^{11}\,{\cal V}[{\rm cm}^3] \,L_0[10^6 {\rm cm}^{-2}] \,  
\sqrt{\frac{T[{\rm mK}]}{\ln [n_c^*]}}\,,
\label{Kelvincontr}
\end{eqnarray} 
the second line giving the absolute magnitude of the specific heat
expressed in terms 
of the temperatures, line densities, and sample dimensions of 
a typical experiment in $^4\!$He.
There is no dependence on $\xi_c$ (save for the very weak one contained in 
$\ln [n_c^*]$), 
as it should be for the large wavelength 
oscillation modes of a vortex line.  
\vspace*{2em}
\begin{center}
\begin{figure}[b]
\psfrag{T[mK]}{$T$\,[mK]}
\psfrag{C/sqrtT}{\large
$\frac {C/k_B}{{\cal V}[\mu {\rm m}^3]\sqrt{T[{\rm mK}]}}$}
\psfrag{L1}{$10^6$\,cm$^{-2}$}
\psfrag{L2}{$10^7$\,cm$^{-2}$}
\psfrag{L3}{${\displaystyle\frac{L_0}{\sqrt{\ln[n_c^*]}}}=2.5\cdot 10^7$\,cm$^{-2}$}
{\includegraphics[width=0.48\textwidth]{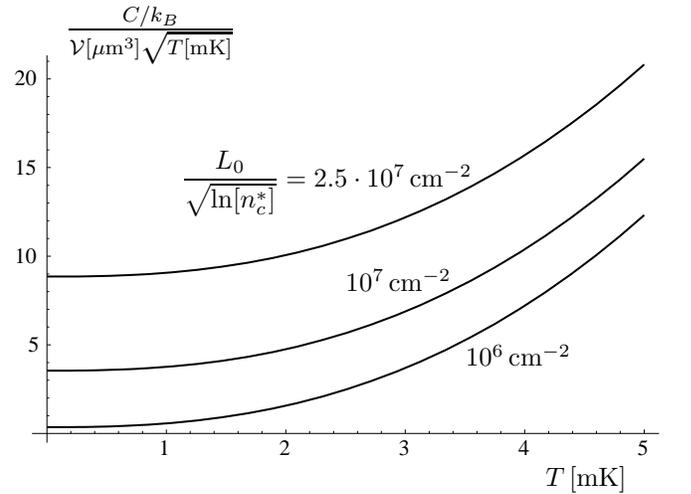}}
\vspace*{1em}
\caption{\label{C} Low temperature asymptotics of the total specific heat 
per volume ${\cal V}={\cal V}
[\mu {\rm m }^3]$, divided by $\sqrt T=\sqrt{T[{\rm mK}]}$,   
for three different values of 
$L_0/\sqrt{\ln[n_c^*]}$ (the speed of sound $c_s= 240$ m/sec).
}
\end{figure}\vspace*{0.5em}
\end{center}
The ratio of the specific heats (\ref{CL}) and (\ref{Cph}) is now given by  
\begin{eqnarray}
\frac{C_{\rm L}}{C_{\rm ph}}& = &
\frac{15\left(\hbar c_s\right)^3 L_0}
{4\pi ^2 \xi_c \left( k_B T\right)^3}\,\, f
\left(\frac{k_B T}{\hbar\omega_c}\right)\nonumber\\
& = & 
\frac{0.74}{\sqrt{\ln [n^*_{c}]}}
\frac1{\sqrt{\Gamma}} 
\left(\frac{\hbar}{k_B T}\right)^{5/2} c_s^3 L_0\,.
\label{ratioExplicit}
\end{eqnarray}  
We may  express this ratio in units of relevance 
for turbulence experiments in superfluid $^4\!$He:  
\begin{eqnarray}
\frac{C_{\rm L}}{C_{\rm ph}}& = &
\frac{1.67}{\sqrt{\ln[n_c^*]}} \,\frac{c^3_s[2.4\cdot 10^4\,{\rm cm/sec}]}{
T^{5/2}[{\rm mK}]}\,
L_0[10^6 {\rm cm}^{-2}]\,. \label{numbers}
\end{eqnarray} 
The contribution of line oscillations to the specific heat thus becomes 
noticeable, using the indicated line density of $L_0=10^6\,{\rm cm}^{-2}$,  
for temperatures which are of the order mK. 
The total specific heat $C=C_{\rm ph}+ C_{\rm L}$ per volume $\cal V$, 
divided by $\sqrt{T}$, is displayed in Fig. \ref{C}. 
The extrapolation of $C/\sqrt T$  
to $T=0$ yields the smoothed line density, up to the very weak 
(square root of) logarithmic 
cutoff dependence on the largest possible 
number of waves on the filaments (the ultraviolet cutoff mode number), 
{\it i.e.}, it gives the quantity  $L_0/\sqrt{\ln[n_c^*]}$. 

In higher temperature superfluid turbulence experiments, 
values of $L\sim 10^8\, {\rm cm}^{-2}$ have been reported \cite{stalp}. 
The number for $L_0$ put into the 
estimates (\ref{Kelvincontr}) and (\ref{numbers}) 
above is thus rather conservative.  
The ultimate limit of achievable line densities in superfluid turbulence, and
in particular the {\em absolute} 
line densities which are reached in turbulence 
experiments at mK temperatures, 
are as yet not clear \cite{hendry2000}. 
Because the conventional second sound 
technique to detect vortices \cite{vinen57}, 
fails at these temperatures, one has to look for
different means to measure the vortex density. One conceivable possibility, 
using a plot like that shown in Fig. \ref{C}, is to measure the specific heat
of the developed turbulent state in the mK range,  divide by $\sqrt T$, 
and extract the (smoothed)  
vortex density from an extra\-polation of the resulting asymptote 
to absolute zero. It should of course be mentioned 
that an experimental determination of specific heats as small as 
those estimated in equation (\ref{Kelvincontr})
is a less than trivial affair
(though not entirely unrealistic).  
However, even if a direct determination 
of the specific heat of line oscillations $C_{\rm L}$ will prove to be
difficult, this quantity is 
of relevance for the (thermo-)dynamical
behaviour of the superfluid at the lowest temperatures. 

In the present context, it is also of interest to note 
that in the numerical simulation work of 
Nore {\it et al.} \cite{noreabidbrachet97}, quite large line densities 
of order $L_0 \sim 10^{10}\cdots 10^{11} {\rm cm}^{-2}$ 
have been assumed. 
The resulting crossover temperatures, for which 
$C_{\rm ph} = C_{\rm L}$, are consequently 
in these simulations already of order 100 mK. However, 
as already alluded to above, it is likely that 
for very large 
densities, modifications of the Kelvin spectrum,  
by mutual induction of the vortex filaments, 
have to be taken into account.

According to (\ref{ratioExplicit}), the ratio of the contributions of 
Kelvin line oscillations and phonons to 
the specific heat scales with $L_0\,T^{-5/2}$, 
so that simultaneously relatively 
low temperatures and high line densities are
necessary to observe the influence of the specific heat of the Kelvin modes
upon the thermodynamical behaviour of the turbulent superfluid. 
From equation (\ref{numbers}), we can 
conclude that the required values are not unreasonably large, and that 
the contribution of the Kelvin 
oscillations in the vortex tangle 
should indeed be 
observable. 

The author 
acknowledges financial support 
by the DFG (FI 690/1-1).

\end{document}